\documentclass[doublecol]{epl2} 

\title{An alternative {scenario} for the formation of specialized protein nano-domains (cluster phases) in biomembranes}
\shorttitle{Formation of specialized protein nano-domains in biomembranes} 
\author{Nicolas Destainville\inst{1,2}}
\shortauthor{N. Destainville}

\institute{                    
  \inst{1} Universit\'e de Toulouse; UPS; Laboratoire de
Physique Th\'eorique (IRSAMC) - F-31062 Toulouse, France\\
  \inst{2} CNRS; LPT (IRSAMC) - F-31062 Toulouse, France}

\abstract{
We discuss a realistic {scenario}, accounting for the existence of sub-micro\-metric protein domains in cell membranes. At the biological level, such membrane domains have been shown to be specialized, in order to perform a determined biological task, in the sense that they gather one or a few protein species out of the hundreds of different ones that a cell membrane may contain. By analyzing the balance between mixing entropy and protein affinities, we propose that such protein sorting in distinct domains can be explained without appealing to pre-existing lipidic micro-phase separations, as in the lipid raft scenario. We show that the proposed scenario is compatible with known physical interactions between membrane proteins, even if thousands of different species coexist.}

\begin{document}

\maketitle

\section{Introduction}

Membrane functional organization is a ubiquitous issue in cell biophysics~\cite{Lenne09,Jacobson07}. It has become consensual that membrane constituents, predominantly lipids and proteins, adopt a non-random, heterogeneous organization~\cite{Engelman05}. Lateral segregation is now accepted as a fundamental requirement for membrane biological functions~\cite{Simons97,Davare01,Laporte01,Rozenfeld10}. The immense variety of lipids and proteins in a single biomembrane (several hundred different species) leads to a large variety of interactions between them. These interactions have been demonstrated to favor the formation of membrane domains, whose size ranges from few nanometers to microns. Domains can be induced by lipid-lipid interactions (from which the concept of ``raft'' emerged~\cite{Simons97,Jacobson07,Pike06}), lipid-protein interactions~\cite{Gil97,Gil98,Sprong01,Poveda08} or protein-protein interactions~\cite{Daumas03,Sieber07,Destain08,Foret08,Gurry09}. Understanding the role of these domains and their physico-chemical origin remains a key problem in cell biology.

The physical scenario investigated in this work belongs to the protein-protein category above (even though lipids do play a role because they indirectly participate in effective inter-protein interactions~\cite{Gil97,Gil98,Poveda08,DeMeyer08,West09}). The proposal that protein-protein interactions can drive the formation of domains independently of a lipidic micro-phase separation has recently been advanced by several research groups~\cite{Daumas03,Sieber07,Destain08,Foret08,Gurry09}. Statistical mechanics arguments have been proposed, relying on the same global mechanism: while short-range attraction favors condensation of membrane proteins in a dense phase, some weaker repulsion at longer-range, the origin of which is still debated, prevents a complete phase separation. The resulting phase at equilibrium is called a ``cluster phase''~\cite{Destain08}, by analogy with similar phases in soft condensed matter~\cite{Stradner}.

\begin{figure}
\begin{center}
\includegraphics*[width=6.5cm]{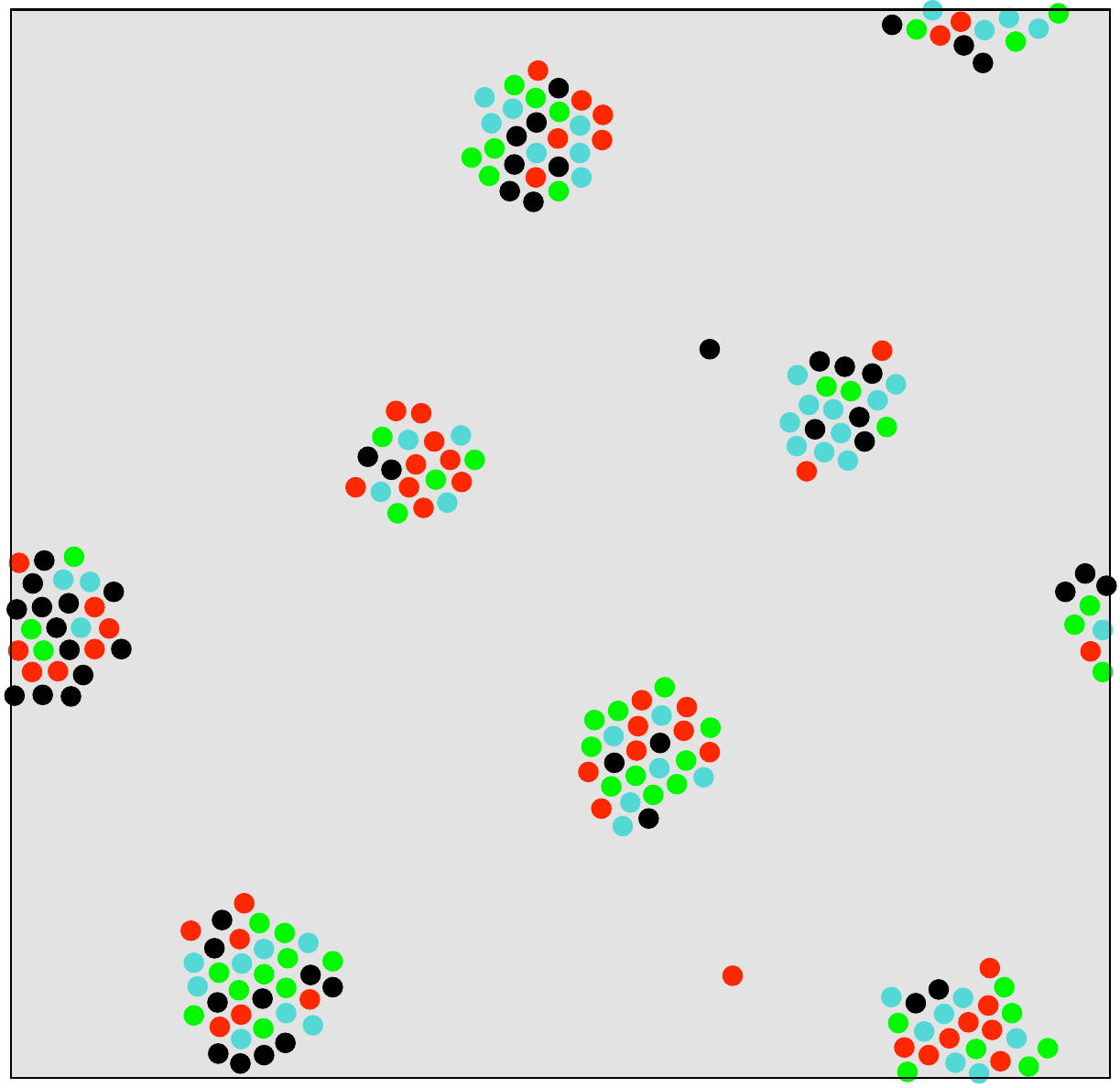}

\includegraphics*[width=6.5cm]{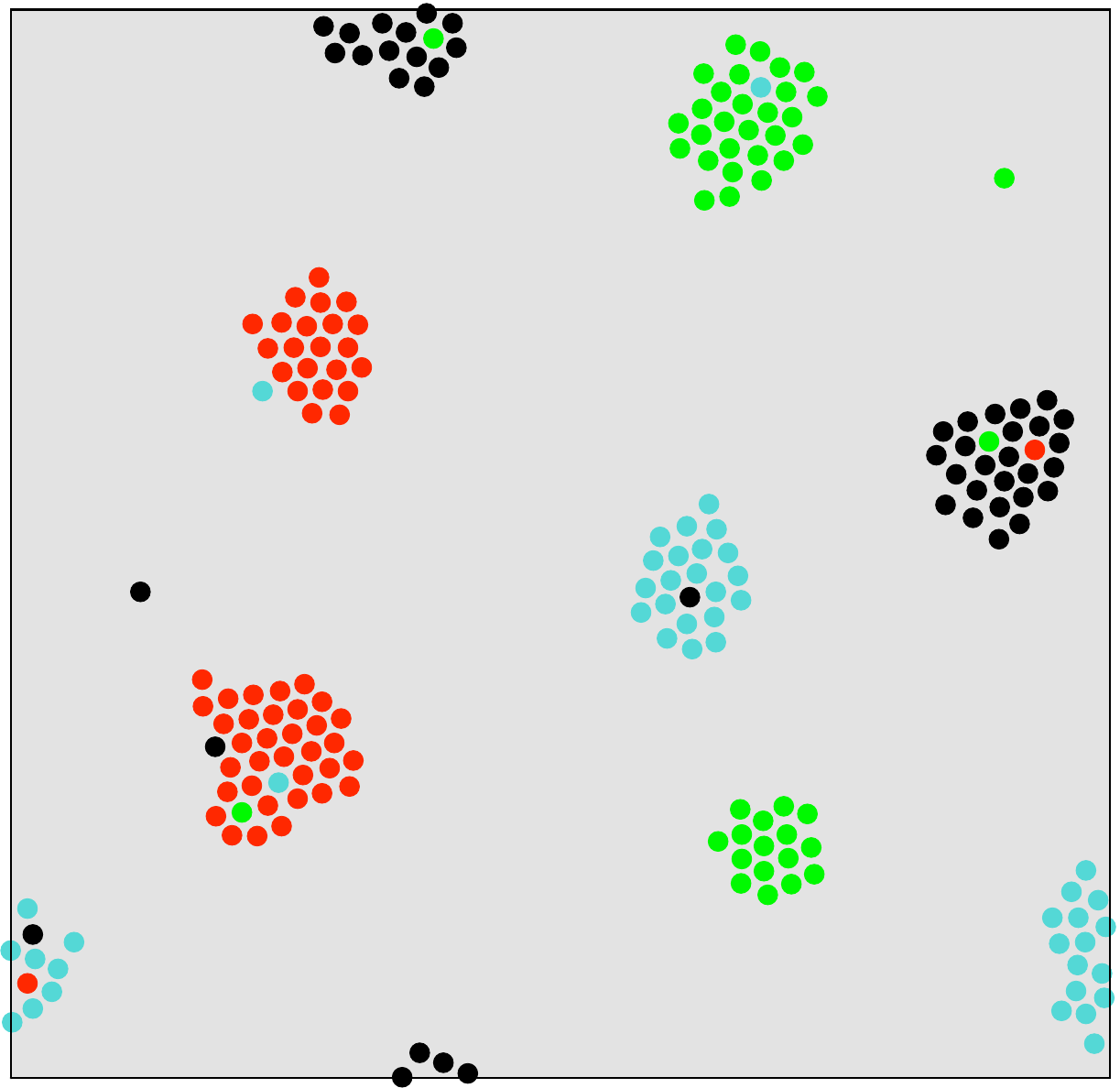}
\end{center}
\caption{Two {schematized views} of membrane cluster phases, with $q=4$ different protein families  (different colors). Top: if proteins of different families have a sufficiently high affinity, they mix together in clusters. Bottom: in the opposite case, clusters are mono-colored, i.e. they are highly concentrated in proteins of one family because the ensuing energy gain is larger than the related entropic cost. This phase-separation leads to cluster specialization because proteins that are destined to co-localize in same clusters indeed do so, spontaneously.}
\label{Clusters}
\end{figure}

However, these studies consider the clusterization of a {\em single} species of proteins, without taking into account the hundreds of different species in the cell membrane with which it coexists. In particular, they do not explain why these other species are excluded from the clusters under consideration. A priori, mixing entropy would stabilize multiple-species clusters (Fig.~\ref{Clusters}). Understanding the specialization of clusters remains a matter in debate of cell biology. Experiments demonstrate the existence of finite-size protein domains gathering a single or few species. They are believed to have a precise biological function, e.g. (high-fidelity) signaling~\cite{Davare01,Prior01B,Daumas03,Park05,Abankwa07,Prior07}, cell adhesion and motility~\cite{Espenel08}, immune response~\cite{DeBakker07A}, endocytosis~\cite{Frick07}, exocytosis and membrane fusion~\cite{Sieber07,Hess07}. 

The present work proposes to solve this issue by coupling the previous cluster-phase model~\cite{Destain08,Foret08}~-- accounting for a finite domain size~-- and a Flory-Huggins theory~\cite{DeGennes}~-- taking into account the great variety of membrane proteins. Proteins are classified into families so that the same-family proteins, which are not necessarily identical, are those that tend to co-localize in same domains to form functional platforms, e.g. a G-protein-coupled receptor and associated G proteins and effectors~\cite{Davare01,Laporte01,Abankwa07}. The references~\cite{Prior01B,Daumas03,Park05,Prior07,Espenel08,DeBakker07A,Frick07,Sieber07,Hess07,Gurry09} provide other examples. The model assumes that the same-family proteins have a higher ``affinity'' at contact than proteins in distinct families, i.e. contact between proteins in the same family is energetically favored. This difference of affinity is measured by a parameter $\chi>0$, reminiscent of a Flory interaction parameter. 

We prove that if $\chi$ is smaller than a critical value $\chi^c$, clusters mix protein families because entropy dominates. If $\chi>\chi^c$, clusters demix and are essentially composed of a single family, the different families being segregated between different specialized clusters (Fig.~\ref{Clusters}). Furthermore, $\chi^c$ is small even if the number $q$ of different families is very large (up to $10^4$) because $\chi^c \propto  k_B T \ln q$, where $k_B T$ is the thermal energy. The main result of the present work is that a contact energy difference as small as 2 to 3 $k_B T$ suffices to favor a demixed cluster phase where clusters are specialized because they gather specific proteins. We show that such contact-energy differences are indeed compatible with known interactions between membrane proteins.

\section{Description of the model}

We consider $N$ {isotropic (or weakly anisotropic)} proteins interacting in a 2D medium that accounts for the lipid ``sea'' in which proteins reside. Even though they are not explicitly taken into account, lipids are responsible for effective interactions between proteins, of elastic or entropic nature (see below). We assume that the total interaction free energy of a protein configuration is the sum of pairwise potentials,
\begin{equation}
U_N({\bf r}_1,\ldots,{\bf r}_N) = \sum_{1\leq i<j\leq N} u_{ij}({\bf r}_i-{\bf r}_j),
\label{U}
\end{equation}
where ${\bf r}_i$ is the position of protein $i$. {$U_N$ does not depend on the protein orientations because we assume that they are isotropic or that anisotropy is sufficiently weak (see Refs.~\cite{Dommersnes99,Chou01} for the quantification of anisotropy effects)}~; $u_{ij}$ depends on $i$ and $j$, because proteins can be of different nature~\footnote{Note that in this paper, all energies and thermodynamic potentials are implicitly expressed in units of $k_BT$.}. Assuming a pairwise interaction is certainly an approximation since many-body effects are known to exist in this context~\cite{Dommersnes99,Kim99,Fournier03,Brannigan07}. Thus the present calculation is only a first step towards the full solution. {We shall return to this point in the Discussion section.}

Following previous works {(see \cite{Sieber07,Destain08,Gurry09} and references therein)}, we assume that the potentials $u$ are attractive at short range ($R < 1$~nm, i.e. roughly speaking ``at contact'') and weakly repulsive at intermediate range ($R > 10$~nm). The repulsion has been given several explanations~\cite{Sieber07,Destain08,Gurry09,McConnell88} which we do not intend to discuss here (see Discussion).  Indeed, in the ``liquid-droplet'' formalism adopted below, the precise shape of the potential does not qualitatively affect the results, in agreement with numerical work~\cite{Destain08}. 
Note that even a weak (a fraction of $k_BT$) intermediate-range protein-protein repulsion  suffices to promote clusters~\cite{Destain08}.

As compared to these previous studies, we introduce a new ingredient in our model: the dependence of the short-range part of $u_{ij}$ on protein families $i$ and $j$. 
Origins  of the energy modulation at contact are manifold: direct electrostatic and polar interactions, even though screened in physiological conditions, play a role at short range; hydrophobic-mismatch interactions depend on transmembrane-protein hydrophobic thicknesses~\cite{Dan93,Brannigan07,Schmidt08}; in membrane with different lipid species, proteins recruit lipids in their immediate vicinity for which they have a higher affinity, because of electrostatic and hydro\-gen-bond interactions between lipids and polar or charged amino acids of proteins. These lipid annuli, which ``wet'' the proteins, are also responsible for effective attractive forces depending on the protein nature~\cite{Poveda08,Gil98}. 
 
{\section{Results}

We first focus on the case where $u_{ij}$ is independent of $i$ and $j$ (identical particles). We recall briefly the formalism and notations introduced in Ref.~\cite{Foret08} (see also \cite{Mitchell81}). The canonical partition function for identical indistinguishable particles is
\begin{equation}
\label{Z}
Z(N) = \frac{\Lambda^{-dN}}{N!} \int_{V^N} {\rm d}{\bf r}_1\ldots {\rm
d}{\bf r}_N\ e^{-U_N({\bf r}_1,\ldots,{\bf r}_N)},
\end{equation}
where the length 
$\Lambda$ making $Z$ dimensionless is set to be the particle diameter~\cite{Foret08}. $V$ is the $d$-dimensional volume where particles evolve. Speaking of membrane nano-domains assumes that one is able to define regions $V_k$ of $V$ that partition the $N$ particles into disjoint clusters: there are $N_1$ monomers, $N_2$ dimers, and so forth, so that $N = \sum k N_k$. Each $k$-particle cluster (or $k$-mer) dwells in a distinct region $V_k$. Then neglecting interactions between the different clusters~\cite{McConnell88,Mitchell81}, the integral in Eq.~(\ref{Z}) can be written as
a product of integrals over the $V_k$. After simple algebraic manipulations~\cite{Foret08}, 
\begin{equation}
\label{Z2}
Z (N) =  \sum_{\{N_k\}} \prod_k 
\frac{1}{N_k!}\left(V\Lambda^{-d} e^{-F(k)}\right)^{N_k},
\end{equation}
where we have introduced the free energy of a $k$-cluster, $F(k)$: $F(1)\equiv0$ for monomers and for $k>1$,
\begin{eqnarray}
\label{Ik}
F(k) & = & -\ln \left\{\frac{\Lambda^{d(1-k)}}{k!} 
\int_{V_k} {\rm d}{\bf r}_1\ldots{\rm d}{\bf r}_{k-1}\
 e^{-U_k}\right\} \\
  & \simeq & -f k +  \gamma k^{\frac{d-1}{d}} + \sigma k^{\alpha}.
  \label{approxF}
\end{eqnarray}
The last approximation, valid when $k \gg 1$, is at the core of the analytical calculations in Ref.~\cite{Foret08} and will also be adopted in the present work~\footnote{Note that as compared to this expression, $k$ was replaced by $k-1$ in Ref.~\cite{Foret08} for technical reasons, which has no consequences in the large $k$ limit of our interest, because it amounts to corrections of the order of $1/k$. The present formulation will be more easily generalized below.}.  Equation (\ref{approxF}) is a generalization of the liquid-droplet model, to which a repulsion term is added (see Ref.~\cite{Foret08} for further details): (i) the first term accounts for the short-range attraction between proteins and $-f<0$ is a bulk free energy per particle; (ii) the second one represents the free-energy cost of the free interface between the cluster and the surrounding fluid and $\gamma>0$ is a line tension. In dimension $d=2$, this term scales as $\sqrt{k}$, which 
means that protein domains are assumed to have a disk shape~\cite{Destain08}. Other shapes, such as stripes, can exist at very high protein concentrations that are not considered in the present work~\cite{Sear99}; (iii) the last term, where $\alpha>1$, takes into account the weak longer-range repulsion, the strength of which is measured by $\sigma>0$. Since it dominates at large $k$, it renders too large clusters unstable and is responsible for their finite size at equilibrium. Note that the exact form of $F$ for small multimers is irrelevant because they are virtually nonexistent in the regime of parameters at hand~\cite{Foret08}.

In practice, we are interested in cases where cluster sizes range from a few dozen to a few thousand particles. This implies that $\sigma<1$ (weak intermediate-range repulsion) and $f,\gamma \gg1$ (moderate short-range attraction)~\cite{Foret08}. For simplicity, we also set $\alpha=3/2$, and $d=2$ because a membrane is two-dimensional. However the present derivations could be extended to any $d$ and $\alpha$. The mean volume 
fraction of $k$-clusters, $c_k$, derives from $Z(N)$ after introducing the chemical potential $\mu$ and switching to the grand-canonical ensemble: 
$c_k\equiv\langle N_k \rangle\Lambda^d/V=\exp[{\mu k-F(k)}]=\exp[{-G(k;\mu)}]$,
where  $G(k;\mu)=F(k)-\mu k = -(f+\mu)k +  \gamma k^{1/2} + \sigma k^{3/2}$
is the grand potential of a $k$-cluster~\cite{Foret08}. The expected value of the protein volume fraction, $\phi\equiv \langle N \rangle \Lambda^d/V$, sets the chemical potential $\mu$ through the condition
$\phi=\sum_{k=1}^{\infty} k c_k$.  At low volume fraction $\phi$, $\mu=\ln c_1<\ln \phi$  takes large negative values and $G(k)$ increases monotonously with $k$:  $c_k$ is maximal at $k=1$ and decreases exponentially with $k$, with typical width $|\mu|^{-1}$. The system is essentially composed of monomers. Above a critical concentration $\phi^c$, the chemical potential $\mu>\mu^c$ is such that $G(k)$ 
has two local minima: one at $k=1$ and one at $k^*\gg1$. The proteins are partitioned between a gas of monomers and clusters of typical aggregation number $k^*$.  At the critical potential $\mu^c$, $G(k;\mu^c)$ has an inflexion point at $k=k^*=k^c$. Thus $\mu^c$ and $k^c$  satisfy $\partial_k G(k^c;\mu^c)=\partial^2_k G(k^c;\mu^c)=0$.  Consequently $\mu^c =-f + \sqrt{3\sigma\gamma}$ and $k^c=1+\gamma/(3\sigma) \gg 1$.

In the general case, where the system contains $q$ families of proteins, one has to introduce new definitions. $M_K$ is  the number of proteins of family $K$ and $N=\sum M_K$. If two particles of families $K$ and $K'$ are adjacent in the same cluster, their binding energy at contact is denoted by $\epsilon_{K,K'}$. Due to thermal agitation, an entropic contribution must be added to this energy, leading to an average free energy per bond $\varphi_{K,K'}\approx\epsilon_{K,K'}$~\cite{Foret08}. This free energy can be embodied in the Flory interaction parameter $\chi_{K,K'}=\nu \varphi_{K,K'}<0$~\cite{DeGennes}, where $\nu$ is the average number of particle neighbors in a cluster bulk ($\nu=6$ in the present case of 2D dense clusters). If $x_K\equiv M_K/N$, the free energy per particle of a homogeneous mixture of $q$ families reads~\cite{DeGennes}:
\begin{equation}
\frac{F}{N}=\sum_{K=1}^q x_K \ln x_K + \frac12  \sum_{K,K'=1}^q \chi_{K,K'} x_K x_{K'}.
\end{equation}

Before tackling the general case with $q$ families of proteins, we first focus on $q=2$, where $A$- and $B$-families coexist. To simplify the discussion, we assume that $\chi_{AA}=\chi_{BB}$ and define $\chi = \chi_{AB}-\chi_{AA}>0$. The canonical partition function becomes $Z(M_A,M_B)$ and Eq.~(\ref{Z2}) must be adapted to this new situation: if $N_{k_A,k_B}$ is the number of clusters containing $k_A$ $A$-proteins and $k_B$ $B$-proteins, then
\begin{equation}
\label{Z:AB}
Z =  \sum_{\{N_{k_A,k_B}\}} \prod_{k_A,k_B} 
\frac{\left(V\Lambda^{-d} e^{-F(k_A,k_B)}\right)^{N_{k_A,k_B}}}{N_{k_A,k_B}!}.
\end{equation}
Here $F(1,0)=F(0,1)=0$ for monomers and 
\begin{equation}
\label{F:kA:kB}
F(k_A,k_B) =  -\ln \left\{\frac{\Lambda^{d(1-k)}}{k_A! \; k_B!} 
\int_{V_k} \prod_{l=1}^{k-1} {\rm d}{\bf r}_l \
 e^{-U_k}\right\} 
\end{equation}
if $k \equiv k_A+k_B \gg 1$. Simplifying this expression to a liquid-droplet-like form [Eq.~(\ref{approxF})] requires to discuss further the internal organization of clusters. Two main scenarios must be envisaged~\cite{Christensen95}: either $A$- and $B$-proteins are demixed in a same $k$-cluster, with an interface between both phases, or they are homogeneously mixed. In contrast to previous studies where clusters are tackled in a canonical formalism because they cannot exchange material with their environment~\cite{Christensen95}, the demixed case is always unfavorable in the present case for the following reason: it implies an energetic cost proportional to $\sqrt{k}$ per cluster due to the interface between $A$- and $B$-rich phases. Thus replacing an assembly of demixed clusters by an assembly of $A$- and $B$-rich ones, of mixed type, is always favorable, because the ensuing entropic cost is of order 1 per cluster, thus $\ll \sqrt{k}$ . Therefore we only consider mixed clusters below. If one defines the $A$-protein fraction of a $k$-cluster, $x=k_A/k$, the Stirling formula leads to
\begin{eqnarray}
\label{F:kA:kBbis} 
F(k_A,k_B) & = & k[x\ln x+(1-x)\ln(1-x)]   \\ \nonumber
 & - & \ln \left\{\frac{\Lambda^{d(1-k)}}{k!} 
\int_{V_k} {\rm d}{\bf r}_1\ldots{\rm d}{\bf r}_{k-1}\
 e^{-U_k}\right\}
\end{eqnarray}
for $k \gg 1$. The first term is the mixing entropy. The second one can now be written in a liquid-droplet-like form and Eq.~(\ref{F:kA:kB}) becomes 
\begin{equation}
F(k_A,k_B) = -f k +  \gamma k^{1/2} + \sigma k^{3/2} +  k h_2(x)
\label{FkAkB}
\end{equation} 
where $-f=\chi_{AA}/2$ and 
\begin{equation}
h_2(x)\equiv \chi x(1-x) + x\ln x+(1-x)\ln(1-x).
\end{equation}
 This expression of $h_2$ is only a mean-field approximation but it will become exact in the large $q$ limit of interest below~\cite{Mittag74}. Furthermore, we have chosen a line tension $\gamma$ independent of $x$ and this is also an approximation. We shall return to this point below.  We now introduce two chemical potentials $\mu_A$ and $\mu_B$ and
$G(k_A,k_B;\mu_A,\mu_B) = F(k_A,k_B)-\mu_A k_A -\mu_B k_B$.
Writing $k_A=xk$ and $k_B=(1-x)k$, minimizing $G$ with respect to $k_A$ and $k_B$ amounts to minimizing it with respect to $k$ and $x$. Beginning with $x$, we remark that $h_2(x)-\mu_A x -\mu_B (1-x)$ is the mean-field free energy of the Ising model in a magnetic field $(\mu_B-\mu_A)/2$. Here we are interested in the situation where no protein family dominates, i.e. where $M_A \approx M_B$ or  $\mu_A \simeq \mu_B$. Thus we focus on the symmetric case $\mu_A = \mu_B \equiv \mu$ (or $M_A = M_B$). Then the mean-field Ising model in a vanishing magnetic field presents a second-order transition at $\chi^c=2$. For a weak difference in protein affinities ($\chi < \chi^c$), $x_\chi=1/2$ is the most probable value and clusters typically contain an equal number of $A$- and $B$-proteins. By contrast, if $\chi > \chi^c$, $h_2(x)$ has two non-trivial minima at $x_\chi$ and $1-x_\chi$. When $\chi$ grows beyond $\chi_c$, clusters are essentially mono-colored. Once  $x_\chi$ is determined, $G$ becomes a function of $k$:
\begin{equation}
G(k;\mu)=-[f+\mu-h_2(x_\chi)]k +  \gamma k^{1/2} + \sigma k^{3/2}.
\label{Renorm:G}
\end{equation}
Therefore the problem of minimization of $G$ with respect to $k$ now amounts to a one-family problem as considered previously, with a renormalized $f$:
$f_{\rm r} = f-h_2(x_\chi)$ and thus a renormalized $\mu^c$.
{Note that if $\chi=0$, then $x_\chi=1/2$ and $f_{\rm r}=f+\ln2$. Indeed
in the two-color case, the canonical partition function $Z$ is normalized by
$M_A!\;M_B!$ instead of $N!$ because same-color particles are indistinguishable. Thus the entropy is increased by $\ln 2$ per particle
as compared to the one-color case (when $M_A=M_B$).} 

The generalization to any $q$ is straightforward: $F$ and $G$ become functions of $k$ and of the number fractions $x_K=k_K/k$ ($K=1,\ldots,q$). We focus again on the most symmetric case: $\chi_{K,K}=\chi_0$ is independent of $K$, $\chi_{K,K'}=\chi_{\rm m}$ if $K \neq K'$ and $M_K=N/q$. We define $\chi=\chi_{\rm m}-\chi_0>0$ and 
\begin{equation}
h_q(x_1,x_2,\ldots,x_q) = \chi \sum_{K<K'} x_K x_{K'} + \sum_{K=1}^q x_K \ln x_K,
\label{hq}
\end{equation}
the mean-field free energy of the $q$-state Potts model~\cite{DeGennes}, exact in the large $q$ limit~\cite{Mittag74}. The transition is first-order when $q \geq 3$. On gets 
{
\begin{equation}
\chi^c= 2\frac{q-1}{q-2}\ln(q-1) \simeq 2 \ln q.
\label{Quicest}
\end{equation}}
At $\chi^c$, the majority number fraction is $x^c = 1-1/q$, {other colors have identical fractions 
$x'=1/[q(q-1)]$,} and $h_q^c \simeq -1/q$. Again, clusters are essentially mono-colored at the transition and beyond{, while colors are mixed if $\chi<\chi^c$. Figure~\ref{Diag} shows the typical phase diagram derived from Eq.~(\ref{Quicest}), and from the fact that $\mu^c$ is renormalized: $\mu^c_{\rm r}=f_{\rm r} + \sqrt{3\sigma\gamma}$, with 
$f_{\rm r} = f - h_q$.} {Note that if $\chi=0$, $f_{\rm r} = f + \ln q$ for the same reason as above.} {It must be emphasized that in this phase diagram the ``transitions'' as determined above are not true thermodynamic transitions because the finiteness of clusters smoothes transitions~\cite{Ruckenstein75}. More technically, the system grand potential $J$ (still in units of $k_{\rm B}T$) can be calculated:
\begin{equation}
\frac{J}{V} = -\sum_{k_1,\ldots,k_K} c_{k_1,\ldots,k_K}
\label{fractions}
\end{equation}
where $c_{k_1,\ldots,k_K}=\exp[-G(k_1,\ldots,k_q;\mu_1,\ldots,\mu_q)]$ is the mean volume fraction of $(k_1,\ldots,k_q)$-clusters~\footnote{Eq.~(\ref{fractions}) implies that $PV=-J=N_{\rm cl} k_{\rm B}T$, where $P$ is the $d$-dimensional pressure and $N_{\rm cl}\equiv V \sum c_{k_1,\ldots,k_K} $ is the total number of clusters. Thus the system behaves like an ideal gaz of clusters.}. 
The fact that $G$ is dominated at large $k$ by $\sigma k^\alpha$ with $\sigma>0$ ensures the uniform convergence of the series and of all its derivatives, and thus prevents any singularity.}

Note however that the real critical value differs from above when $q>k$, because the $x_K$ do not take continuous but discrete values, $x_K=k_K/k$ in a $k$-cluster ($k_K$ an integer). Thus no more than $k$ variables $x_K$ can be simultaneously non-vanishing and $h_q$ is to be minimized on this subset of $\mathbf{R}^d$, which amounts to a minimization of $h_k$. Thus $\chi^c \simeq 2 \ln k$ and $x^c \simeq 1-1/k$. Generally speaking, if $k^*$ still denotes the typical cluster size, then $\chi^c \simeq 2 \ln [\min(k^*,q)]$. 
{In a similar way, in the phase diagram of Fig.~\ref{Diag}, $q$ should be replaced by $\min(k^*,q)$.}

\section{Discussion}

By appealing only to generic {two-body} protein-protein interactions, we have shown that protein nano-domains may spontaneously form in biomembranes at equilibrium.
One of the goals of the present approach is to account for the finite size of membrane nano-domains. Out-of-equilibrium arguments have also been developed to explain this finite size~\cite{Foret05,Turner05}. In this respect, our goal here has not been to discuss the relative merits of equilibrium or out-of-equilibrium approaches, which are certainly complementary,  but to show that the equilibrium approach alone can yield finite-size nano-domains.
{Exploring the additional effects of out-of-equilibrium phenomena in the cluster phase scenario could clarify their role. In particular, taking into account membrane recycling~\cite{Foret05,Turner05} or proteins interacting differently with their partners when in an active state~\cite{Phillips09} should be tractable in the present formalism.}

\begin{figure}
\begin{center}
\includegraphics*[width=0.95\linewidth]{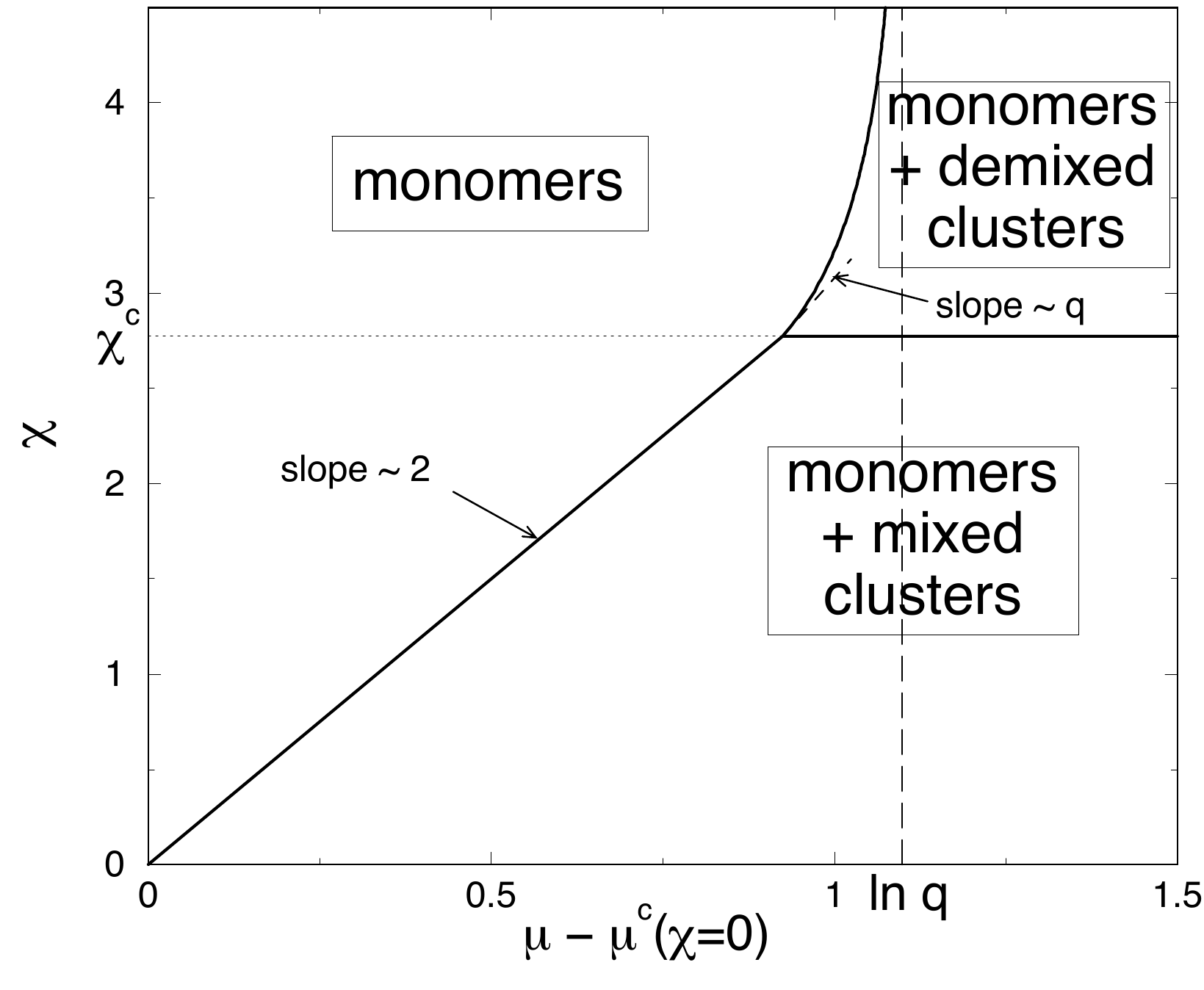}
\end{center}
\caption{{Phase diagram in the $(\mu,\chi)$ plane. It is calculated here in the case $q=3$ but its qualitative aspect is valid for any $q$, with $\chi_c \simeq 2 \ln q$ [Eq.~(\ref{Quicest})]. The lines represent crossovers and not true thermodynamic transitions, as discussed in the text.
}}
\label{Diag}
\end{figure} 
 
Furthermore, we have shown {that, in spite of the associated entropic cost,} it becomes favorable to demix different protein families in distinct domains when $\chi>\chi^c$ with $\chi_c
\simeq 2 \ln [\min (k^*,q)]$. Here $k^*$ is the typical cluster size. Even if $\min(k^*,q) \sim 10^3$ to $10^4$, a realistic upper-bound for a real plasma membrane, $\chi^c \approx 14$ to 18, which means that $|\epsilon_{K,K}-\epsilon_{K,K'}|\simeq |\varphi_{K,K}-\varphi_{K,K'}|  \equiv \chi^c/\nu\approx 2$ to $3 k_BT$. Is it realistic to expect such differences of contact energy between different protein families?  The precise calculation of interaction energies for proteins embedded in a lipidic membrane is a tedious task that can be hardly tackled analytically. By contrast, several numerical studies have estimated the variations of the energy at contact for transmembrane proteins with variable hydrophobic thicknesses~\cite{Brannigan07,Schmidt08,DeMeyer08,West09}. All these studies are consistent with contact energy differences larger than 3 $k_BT$. As far as lipid-mediated interactions are concerned, we are not aware of any quantitative results in the literature. However, given the order of magnitude of this interaction~\cite{Gil97,Gil98}, its amplitude is likely to be modified by one $k_BT$ or more due to interplay between lipid and protein species. Thus typical modulations of $|\epsilon_{K,K}-\epsilon_{K,K'}|$ can easily exceed 2 or 3 $k_BT$. {At the biological level, this result shows that an important concentration of proteins in a biological membrane leads to their gathering and sorting in nano-domains. By contrast, below the critical concentration, they would be distributed randomly on the membrane. To this respect, increasing the concentration strongly facilitates the encounter of proteins having a contact energy slightly lower than the average one, and thus favors biological functions.}

The line tension $\gamma$ measures the free energy cost for a protein to be at the cluster boundary as compared to the bulk. Indeed, a bulk particle has about twice as many neighbors as one at the boundary. Thus 
one would expect $\gamma$ to depend on $x_\chi$ and thus on $\chi$.
But clusters are essentially mono-colored at the transition and beyond, and one expects 
$\gamma$ to depend only weakly on $\chi$ in this region of the phase diagram. 

{To derive the form of the cluster free-energy $F(k_A,k_B)$ in Eq.~(\ref{FkAkB}), or its $q$-color counterpart [see Eq.~(\ref{hq})], we have assumed pairwise interaction potentials in Eq.~(\ref{U}), because inferring $F(k_A,k_B)$ in this particular case is easier~\cite{Foret08}. Real potentials contain many-body contributions that have for example been explored in Refs.~\cite{Dommersnes99,Fournier03,Brannigan07}. In another work, long-range repulsion has been proposed to arise from steric interactions, of many-body origin by nature~\cite{Sieber07}. Even though it is a complex issue, fully taking into account many-body interactions will be necessary to derive properly the effective parameters $f$, $\gamma$, $\sigma$ or $\chi$ from first principles. Numerical simulations, that are out of the scope of the present work, will certainly be required, because, e.g., the full calculation of long-range $N$-body potentials mediated by the elastic membrane requires to compute the inverse and the determinant of a $N$ by $N$ matrix~\cite{Fournier03}. The estimation of many-body effects associated with hydrophobic mismatch also requires intensive simualtions~\cite{Brannigan07}.
Note however that our whole argument relies on the effective free energy $F(k_1,\ldots,k_q)$ and not on the exact shape of the potential $U_N$ from which $F(k_1,\ldots,k_q)$ ensues. }

{It would be interesting to investigate the role of strongly anisotropic inclusions in the future, because in this case, the long-range interaction can become attractive~\cite{Chou01}. We anticipate that the cluster-phase scenario remains valid if only a minority of proteins are strongly anisotropic, that are homogeneously mixed in clusters (we recall that proteins of a same family are not necessarily identical but are classified according to their {\em short-range} affinities~; thus they are miscible in each cluster). Indeed, in this case, the repulsive energy still grows faster than the cluster aggregation number $k$ (i.e. $\alpha>1$ in Eq.~(\ref{approxF})), because the majority of long-range interactions in the cluster remain repulsive. The main ingredient for the existence of stable clusters is preserved~\cite{Destain08}.}

We have also assumed that $\chi_{K,K'}$ can only take two possible values, $\chi_0$ if $K=K'$ or $\chi_{\rm m}$ if $K \neq K'$. The reality is more complex because of the great variety of membrane proteins. Future studies would also have to consider more realistic distributions of $\chi$-parameters around these typical values. The asymmetric case where $M_K$ (or $\mu_K$) depends on $K$ should also be explored. The function $G(\{k_K\};\{\mu_K\})$ will then have to be minimized numerically, providing distributions of the $x_K$ and cluster sizes $k$. However, we anticipate that our conclusions will be qualitatively unchanged since the present approach captures the essential physical ingredient, namely the competition between energy and configurational entropy. 

{
Finally, our predictions ought to be tested experimentally. F\"orster Resonance Energy Transfer (FRET) is an appropriate tool to quantify the distance between tagged bio-molecules~\cite{FRET}. After incorporating different proteins species in giant vesicles, tagged with different fluorophores~\cite{Abankwa07}, it would be possible to test qualitatively the predictions of the present work by playing on physical parameters such as the asymmetry of the proteins or the thickness of the lipidic membrane.  
}
\acknowledgments
I am indebted to Manoel Manghi and Laurence Salom\'e for helpful discussions and comments, and to Revaz Ramazashvili for his kind reading of the manuscript.

\end{document}